\documentclass[12pt]{article}

\DeclareSymbolFont{AMSb}{U}{msb}{m}{n}
\DeclareMathSymbol{\N}{\mathbin}{AMSb}{"4E}
\DeclareMathSymbol{\Z}{\mathbin}{AMSb}{"5A}
\DeclareMathSymbol{\C}{\mathbin}{AMSb}{"43}

\def\sech{{\mathrm{sech}\,}}

\font\fff= eufm10 scaled \magstep 1
\def\?#1?{\hbox{\fff #1}}

\def\ad #1.{{{\rm ad}\,#1}}
\def\X{{\hat X}}
\def\Y{{\hat Y}}
\def\D{{\hat D}}

\def\V{{\hat V}}
\def\W{{\hat W}}
\def\y{{\hat y}}
\def\e{{\bf e}}
\def\1{{\bf 1}}

\def\eqref#1{(\ref{#1})}

\begingroup 
\newtheorem{thm}{Theorem}[section]           

\newtheorem{theorem}[thm]{Theorem}

\newtheorem{notation}{Notation.}

\newtheorem{remark}[thm]{Remark. }             
\newtheorem{example}[thm]{Example}           
\endgroup

\def\qed{\qquad\framebox[7pt] \medskip \noindent}

\def\th{\thinspace}

\def\avg#1.{\langle#1\rangle}
\def\vstrut{\phantom{\biggm|}}

\newenvironment{proof}{{\sl Proof:}\quad}{\hfill{\qed}\\ \noindent}

\renewenvironment{abstract}{\begin{quote}{\bf Abstract.\
}\small}{\end{quote}\bigskip}

\def\vstrut{\phantom{\biggm|}}

\title{Finite-Dimensional Calculus}

\author{Philip Feinsilver\thanks{\leftskip -1cm Department of Mathematics
Southern Illinois University 
 Carbondale, IL. 62901, U.S.A.}
\  and Ren{\'e} Schott\thanks{\leftskip -1cm IECN and LORIA, 
 Universit\'e Henri Poincar\'e-Nancy 1, BP 239, 
54506 Vandoeuvre-l\`es-Nancy, \hbox to 18pt{}France.}} 

\date{}

\begin{document}

\maketitle
\thispagestyle{empty}

\begin{abstract}
We discuss topics related to finite-dimensional calculus in the context of finite-dimensional quantum mechanics.
The truncated Heisenberg-Weyl algebra is called a TAA algebra after Tekin, Aydin, and Arik who formulated it in terms of orthofermions.
It is shown how to use a matrix approach to implement analytic representations of the Heisenberg-Weyl algebra in univariate
and multivariate settings. We provide examples for the univariate case. Krawtchouk polynomials are presented in detail, 
including a review of Krawtchouk polynomials that illustrates some curious properties of the Heisenberg-Weyl algebra, 
as well as presenting an approach to computing Krawtchouk expansions. 
From a mathematical perspective, we are providing indications as to how to implement in finite terms Rota's ``finite operator calculus".

\end{abstract}
 
\vfill
\pagebreak

\section{Introduction}
Ever since Weyl's ``Group Theory and Quantum Mechanics", the question of finite-dimensional representations of the canonical
commutation relations has been of interest. The simplest formulation, two operators $A,B$ on a Hilbert space obeying
$$[A,B]=c\1$$
for a non-zero scalar $c$, 
$\1$ being the identity, is quickly dispatched on a finite-dimensional space $\cal{H}$ by taking traces, the left side yielding zero,
the right giving $c\,\dim \cal{H}$.\bigskip

We illustrate with the interesting phenomenon of ``Krawtchouk ghosts", where the Krawtchouk polynomials, polynomials orthogonal
with respect to a binomial distribution, are an infinite family of polynomials that carry a representation of the HW commutation relations,
but span a finite-dimensional Hilbert space, all but a finite number of the polynomials having zero norm. It turns out that
the Krawtchouk polynomials in each dimension are a basis for corresponding representations of $\?su?(2)$, used ubiquitously in
current work on quantum information. However, they do not appear explicitly yet in current work on quantum information.\bigskip

Mathematical physicists have used Krawtchouk polynomials to develop finite-dimensional quantum mechanics.
See Lorente \cite{Lo1,Lo2} for work related to this approach. Atakishiyev \& Wolf \cite{AW,APW} have used Krawtchouk polynomials
to formulate finite-dimensional wave functions to replace Gaussian packets, e.g. in optics. 
Santhanam \cite{SA} has discussed the difficulties as well as possibilities of finite-dimensional quantum mechanics. 
In other domains, 
Krawtchouk expansions have been used in image analysis showing results superior to many commonly used techniques \cite{Y}.\bigskip

The approach here will show how up to a given order one may
produce representations of the HW algebra on spaces of polynomials that we call ``analytic representations", whereby
corresponding to any function analytic in a neighborhood of the origin in $\C$, one produces a sequence of {\sl canonical
polynomials,\/} that are closely related to many of the polynomial sequences arising in physics, probability theory,
combinatorics, etc. \cite{FS,R,RR}.\bigskip

Another approach to finite-dimensional representations of the canonical commutation relations is via replacing the identity
by an operator $E$ with the defining relations
\begin{equation}\label{HW}
[A,B]=E,\quad [A,E]=[B,E]=0
\end{equation}
whereby $E$ is central. This is illustrative of the inapplicability of Schur's Lemma where a central element must be a multiple of
identity as for finite-dimensional representations of unitary groups. The finite-dimensional representations of 
the relations \eqref{HW} have been found using a diagrammatic approach in \cite{GDF}. 
\bigskip

There are three main aspects of this work:

\begin{enumerate}
\item We make the observation that the TAA algebra is the appropriate formulation for abstracting the structure of the
truncated Heisenberg-Weyl algebra. We mention related work in the area of discrete/finite quantum mechanics.\medskip

\item We show that for analytic representations of the HW algebra, computations using truncated HW can be done
order-by-order numerically, avoiding the necessity for symbolic computations.\medskip

\item We show some properties of Krawtchouk polynomials that are of interest with regards to HW representations.\medskip

\end{enumerate}

The algebra we call here ``TAA algebra'' was presented by Tekin, Aydin, and Arik in \cite{TAA}.
It is generated by an operator $a$ and its adjoint $a^*$. Setting $\nu=a^*a$, the commutation rule defining the algebra is
$[a,\nu]=a$. Taking adjoints gives the complementary rule $[\nu,a^*]=a^*$. In other words, we assume that $\nu$ 
and $a$ generate the two-dimensional Lie algebra of the affine group. Writing it out
\begin{equation}\label{eqTAA}
 aa^*a-a^*aa=a
\end{equation}
we see that the commutation rule $aa^*-a^*a={\1}$ of the Heisenberg-Weyl algebra has been multiplied by $a$ on the right. This modification 
is enough to yield finite-dimensional representations, including the truncated Heisenberg-Weyl algebra (HW-algebra) given by the operators $X=$multiplication by $x$ and
$D=d/dx$ acting on polynomials of a given bounded degree. Writing matrices for these operators we will see that they 
obey equation \eqref{eqTAA}. \bigskip

This article may be thought of as realizing Rota's idea of ``Finite Operator Calculus'' \cite{R,RR} in a truly finite way.
There is work of P.\th R.\th Vein along similar lines \cite{V1,V}.
The main feature here is that the operator calculus
is done on finite-dimensional spaces and can be carried out explicitly using matrices. 
The approach in this paper is based on algebraic
properties of the operators and includes a formulation for the multivariable case. 
The one-variable case is dual to that presented in \cite{FS1}.\bigskip

\section{Orthofermion formulation}
Here we recall the orthofermion approach of \cite{TAA}. 
Start with a set of operators $\{c_1,\ldots,c_p\}$, with $p$ a positive integer and form the star-algebra
generated by the $\{c_i\}$ modulo the following relations
\begin{eqnarray}\label{eqORTHO}
c_ic_j&=&0 \nonumber \\
c_ic_j^*+\delta_{ij}\,\sum_{k=1}^pc_k^*c_k &=& \delta_{ij}\,{\1}
\end{eqnarray}
where $\1$ is the identity operator. Setting $\Pi={\1}-\sum\limits_{k=1}^pc_k^*c_k$ (as in \cite{M}) we can write this last relation as
$$c_ic_j^*=\delta_{ij}\,\Pi$$
where one readily shows from the defining relations (\ref{eqORTHO}) that $\Pi^2=\Pi$, i.e., $\Pi$ is a projection as suggested by the notation.\bigskip

It follows from the defining relations,  that $\Pi c_k=c_k$ and from the second relation of eq. (\ref{eqORTHO}) we have the useful relation
\begin{equation}\label{eqFORM}
 c_ic_j^*c_k=\delta_{ij}\,c_k
\end{equation}

Within the orthofermion algebra, following \cite{TAA}, modifying slightly their formulation, we set
\begin{eqnarray*}
a&=&c_1+\sum_{k=2}^p k\,c_{k-1}^*c_k\\
a^\dagger&=&c_1^*+\sum_{k=2}^p c_k^*c_{k-1}\\
\end{eqnarray*}
Using equation (\ref{eqFORM}), we get
$$aa^\dagger-a^\dagger a={\1}-(p+1)\,c_p^*c_p$$
which then yields the relation corresponding to equation (\ref{eqTAA}) of the TAA algebra.

\section{Calculus with matrices}\label{secCM}

\begin{notation}\rm
We will denote the matrix corresponding to an operator by using a $\hat{}$ symbol. Thus, $\hat X$ is the matrix corresponding to the operator
$X$, etc.
\end{notation}

Restricting the differentiation operator to the finite-dimensional space of polynomials of degree less than or equal to $p$ is no problem.
Use the standard basis $\{1,x,x^2,\ldots,x^p\}$. For $p=4$, we have
$$\D = \pmatrix{0&1&0&0&0 \cr 0&0&2&0&0 \cr 0&0&0&3&0\cr 0&0&0&0&4\cr 0&0&0&0&0}$$
with the extension to general $p$ following the same pattern. However, multiplication by $x$ must be cut off. If we define $X\,x^i=x^{i+1}$ for $i<p$ and
$X\,x^p=0$, we no longer have the relation $DX-XD=\1$. Instead, we have the TAA relation
$$DXD-XDD=D$$
The matrix of $X$ has the form, for $p=4$, 
$$\X = \pmatrix{0&0&0&0&0 \cr 1&0&0&0&0 \cr 0&1&0&0&0\cr 0&0&1&0&0\cr 0&0&0&1&0 \cr}$$
Note that ${\X}^{p+1}=0$.
To keep in line with the powers of $x$, we label the basis elements starting from $0$. So
let $\e_k$ denote the column vector with the only nonzero entry equal to $1$ in the $(k+1)^{\rm st}$ position.
The vacuum state is $\Omega=\e_0$,  satisfying $\D\Omega=0$. And ${\X}^k\Omega=\e_k$, for $1\le k\le p$.
As expected, these are raising and lowering operators satisfying
\begin{displaymath}
\X\e_k=\e_{k+1}\,\theta_{kp}\,,\qquad \D\e_k=k\,\e_{k-1}
\end{displaymath}
where $\theta_{ij}=1$ if $i<j$, zero otherwise.\\

With the inner product $\avg \e_n,\e_m.=\delta_{nm}\,n!\,$, we indeed have $\D^*=\X$.\\

Let $E_{ij}$ denote the standard unit matrices with all but one entry equal to zero, $(E_{ij})_{kl}=\delta_{ik}\delta_{jl}$, $1\le i,j,k,l\le p+1$.
The connection with orthofermions is given by the $(p+1)\times(p+1)$ matrix realization
$$ \hat c_i=E_{1\,i+1}$$
for $1\le i \le p$. 
The orthofermion relations hold and particularly for this realization
$$\hat c_i^*\hat c_j=E_{i+1\,j+1}$$
Note that 
$\hat\Pi=E_{11}$ and that the star-algebra generated by the $\hat c_i$ is the full matrix algebra.\\

As long as $X$ never multiplies the power $x^p$, the matrix implementation agrees with usual calculus. The TAA relation formulates
this algebraically.\\

The following theorem shows that $\D$ and $\X$ not only do not generate a Heisenberg algebra, but, in fact,
are as far as possible from doing so.

\begin{theorem}\label{thmSL}
For $p>0$, let $\D$ and $\X$ be $(p+1)\times (p+1)$ matrices defined by 
$\displaystyle \D=\sum_{k=1}^p k\,E_{k\,k+1}$,
$\displaystyle \X=\sum_{k=1}^{p} E_{k+1\,k}$. 
Then the Lie algebra generated by $\{\X,\D\}$ is $\?sl?(p+1)\vstrut$.
\end{theorem}

\begin{proof}     
For convenience set $n=p+1$. First we have $$H=[\D,\X]=-p\,E_{nn}+\sum_{k=1}^p E_{kk}$$
Set $\xi_1=\X$, $\eta_1=\D$, and $H_1=H$.
For $2\le k\le n$, let $\xi_k=H(\overleftarrow{\ad \X.})^k$, and $\eta_k=(\ad \D.)^kH$, where $(\ad A.)B=[A,B]$ and
$A(\overleftarrow{\ad B.})=[A,B]$. Then it is easily checked by induction that
$$\xi_k= -n\,E_{n\,n-k+1}\qquad \hbox{and}\qquad \eta_k=a_k\,\xi_k^\dagger$$
for nonzero constants $a_k$, the $\dagger$ denoting matrix transpose. 
Thus, we obtain $E_{in}$ and $E_{ni}$ for $1\le i\le p$. 
Noting that $[E_{in},E_{nj}]=E_{ij}$ if $i\ne j$, we have all of the off-diagonal $E$'s.
And 
$$ H_k=[\eta_k,\xi_k]=-na_k\, (E_{n-k+1\,n-k+1}-E_{nn})$$
fill out the Cartan elements of $\?sl?(n)$.
\end{proof}

\subsection{Examples}
Here we look at some important operators.
\begin{example}     \rm
The {\it number operator\/} is $XD$. For $p=4$, we have
$$\X\D= \left( \begin {array}{rrrrr} 0&0&0&0&0\\\noalign{\medskip}0&1&0&0&0
\\\noalign{\medskip}0&0&2&0&0\\\noalign{\medskip}0&0&0&3&0
\\\noalign{\medskip}0&0&0&0&4\end {array} \right)$$
This operator multiplies $\e_n$ by $n$, for $0\le n\le 4$.
In general, we have
$$\X\D=\sum_{n=0}^p n\,E_{n+1\,n+1}$$
which multiplies $\e_n$ by $n$, for $0\le n\le p$.
\end{example}

\begin{example} \rm
The Hermite polynomials, occurring in oscillator wave functions, are eigenfunctions of
the {\it Ornstein-Uhlenbeck operator\/}, $XD-tD^2$, $t>0$, which for $p=4$ takes the form
$$\left( \begin {array}{rrrrr} 0&0&-2t&0&0\\\noalign{\medskip}0&1&0&-6t&0
\\\noalign{\medskip}0&0&2&0&-12t\\\noalign{\medskip}0&0&0&3&0
\\\noalign{\medskip}0&0&0&0&4\end {array} \right)
$$
The eigenvector for each eigenvalue $\lambda=0,1,2,3,4$ gives the coefficients of the corresponding
polynomial $H_\lambda(x,t)$.  The family of polynomials $\{H_\lambda(x,t)\}_{\lambda\in\N}$ provide an orthogonal basis
for $L^2$ with respect to the Gaussian measure with mean zero and variance $t$.
\end{example}

\begin{example} \rm
The {\it translation operator} $T_t=e^{tD}$ acts on functions as $e^{tD}f(x)=f(x+t)$. For $p=4$, 
$$\hat T_t= \left( \begin {array}{rrrrr} 1&t&t^2&t^3&t^4\\\noalign{\medskip}0&1&2t&3t^2&4t^3
\\\noalign{\medskip}0&0&1&3t&6t^2\\\noalign{\medskip}0&0&0&1&4t
\\\noalign{\medskip}0&0&0&0&1\end {array} \right)
$$
generally, with columns given by binomial coefficients times powers of $t$, corresponding to the action $x\to x+t$ on the basis polynomials
$x^j$. The matrix $\hat T_t$ can be computed as the exponential of $t\hat D$ defined as a power series: 
$\1+t\hat D+t^2\hat D^2/2!+\cdots$ .
\end{example}

\begin{example} \rm
The Gegenbauer polynomials satisfy 
$$ [(XD+\alpha)^2-D^2]C_n^\alpha(x)=(n+\alpha)^2\,C_n^\alpha(x)$$
see, e.g. \cite{FF}. Thus we have the {\it Gegenbauer operator,\/} $G_\alpha=(XD+\alpha)^2-D^2$, which for $p=4$ takes the form
$$\hat G_\alpha=\left( \begin {array}{ccccc} {\alpha}^{2}&0&-2&0&0
\\\noalign{\medskip}0& \left( 1+\alpha \right) ^{2}&0&-6&0
\\\noalign{\medskip}0&0& \left( 2+\alpha \right) ^{2}&0&-12
\\\noalign{\medskip}0&0&0& \left( 3+\alpha \right) ^{2}&0
\\\noalign{\medskip}0&0&0&0& \left( 4+\alpha \right) ^{2}\end {array}
 \right)
$$
where the spectrum is evident along the diagonal. Up to order $p$, one obtains the Gegenbauer polynomials with coefficients
given by the eigenvectors of $\hat G_\alpha$.
\end{example}

\subsection{Multivariable calculus with matrices}

Here we extend to $N$ variables. For matrices, $A,B$, the tensor product $A\otimes B$ denotes the
{\sl Kronecker product\/} of the two matrices. That is, if $A$ is $n\times n$, and $B$ is $m\times m$, then $A\otimes B$ is
$nm\times nm$ with entries formed by replacing each entry $a_{ij}$ in $A$ with the block matrix $a_{ij}B$. 
For products of more than two matrices, we conventionally associate to the left.\\

For a fixed $p$, we have $(p+1)\times (p+1)$ matrices $\D$ and $\X$. 
Let $I$ denote the $(p+1)\times (p+1)$ identity matrix. Then we set
\begin{eqnarray*}
\D_j &=& I \otimes I \otimes\cdots\otimes \D \otimes I\cdots \otimes I  \qquad\hbox to 30pt{}(\D \hbox{ in the }j^{\rm th}\hbox{ spot})\\
\X_j &=& I \otimes I \otimes\cdots\otimes \X \otimes I\cdots \otimes I  \qquad\hbox to 30pt{}(\X \hbox{ in the }j^{\rm th}\hbox{ spot})
\end{eqnarray*}

Then $\D_j$ and $\X_j$ will satisfy the TAA relations while $[\D_j,\X_i]=[\X_j,\X_i]=[\D_j,\D_i]=0$ for $i\ne j$.

\section{Analytic representations of the HW-algebra. Canonical polynomials}
Now we would like to discuss {\sl analytic representations\/} of the HW-algebra. These are infinite-dimensional representations
in the sense that they act on a basis for the vector space of polynomials in a given set of variables 
$\{x_1,x_2,\ldots,x_N\}$. Basic to our approach is the use of {\sl canonical variables\/} which are functions of $X$ and $D$ obeying the HW relations
on an infinite-dimensional space, which restricts to the TAA relation on spaces of polynomials in $x$ of a given bounded degree.\\

Let us review the basic construction and notations for the general, multivariable, case.\\

\begin{notation}  \rm
We use the convention of summing over repeated Greek indices, {\sl irrespective of position.}
\end{notation}

Given 
$V\colon\mathbf{C}^N\to\mathbf{C}^N$,
$V(z)=(V_1(z_1,\ldots,z_N),\ldots,V_N(z_1,\ldots,z_N))$ 
holomorphic in a neighborhood of the origin, satisfying $V(0)=0$,
we construct an associated abelian family of dual vector fields.
Corresponding to the operators $X_i$ of multiplication by $x_i$, we have  the partial differentiation
operators, $D_i$.  In this context, a function of $x=(x_1,\ldots,x_N)$, $f(x)$, is
identified with $f(X)1$, the operator of multiplication by $f(X)$
acting on the \emph{vacuum state} $1$, with $D_i1=0$, for all $1\le i\le N$.
We define operators $V(D)=(V_1(D_1,\ldots,D_N),\ldots,V_N(D_1,\ldots,D_N))$.
These are our canonical lowering operators, corresponding to differentiation.\\

Denoting the Jacobian $\displaystyle \left(\frac{\partial V_i}{\partial z_j}\right)$ 
by $V'(z)$, let $W(z)=(V'(z))^{-1}$, be the inverse (matrix inverse) Jacobian. Then the boson commutation relations extend to
$\displaystyle [V_i(D),X_j]=\frac{\partial V_i}{\partial D_j}$. Now define the operators 
\begin{displaymath}
          Y_i=X_\mu W_{\mu i}(D)
\end{displaymath}
These are our canonical raising operators, corresponding to multiplication by $X_i$.  
We have
$$[V_i,Y_j]=\delta_{ij}{\1}$$
Thus, the canonical system of raising and lowering operators is $\{Y_j\}$, $\{V_i\}$, $1\le i,j \le N$. 
The essential feature, which has to be checked, is that, 
$[Y_i,Y_j]=[V_i,V_j]=0$. Notice that exchanging $D$ with $X$ is a formal
Fourier transformation and turns the variables $Y_i$ into
the vector fields $\displaystyle \tilde Y_i=W(x)_{\mu i}\frac{\partial}{\partial x_\mu}\,.$ 
Thus, the $Y_i$ are {\sl dual vector fields\/} \cite{FS}. \\

\begin{notation}
\rm  We complement the standard notations used along with $V$ and $W$, letting
  $U$ denote the inverse function to $V$. I.e., 
$U\circ V=V\circ U=\mathrm{id}$. Explicitly: $U(V(z))=z$.
\end{notation}

Observe that since $W=V'^{-1}$, we have $W(z)=U'(V(z))$. In other words, 
converting from $z$ to $V$ acting on functions of the canonical variables $Y_i$, we have the {\sl recurrence relation}
$$X=Y\,U'(V)^{-1}\ .$$

Using multi-index notation, $n=(n_1,\ldots,n_N)$, $v^n=v_1^{n_1}v_2^{n_2}\cdots v_N^{n_N}$,
the main formula (cf. \cite[p. 185, eq. (1)]{FS}) is
$$ \exp(v_\mu Y_\mu)\,1 = \exp{x_\mu U_\mu(v)}= 
\sum_{n\ge0}{v^n\over n!}\,y_n(x)$$

This expansion defines the {\sl canonical polynomials\/}: $y_n(x)=Y^n\,1$.

\subsection{Canonical Appell systems}
An {\sl Appell system\/}, $\{h_n(x)\}$, in one variable is a system of polynomials providing a basis for the vector space of polynomials
with $\deg h_n=n$, $n=0,1,2\ldots$, such that $Dh_n=nh_{n-1}$. Defining the raising operator $R$ by $Rh_n=h_{n+1}$, we have
$[D,R]=\1$, thus a representation of the HW-algebra. (See \cite[v.\th3, Ch.\th1]{FS2} for further elaborations.)\bigskip

Introducing a Hamiltonian $H(z)$, in this context the only requirement being analyticity in a neighborhood of the origin in $\C^N$,
we have the time-evolution
\begin{equation}\label{cAppell}
\exp{\bigl(-tH(D)\bigr)}\,e^{xU(v)}=e^{xU(v)-tH(U(v))}=\sum_{n\ge0}{v^n\over n!}\,y_n(x,t)
\end{equation}
An Appell system of polynomials has a generating function of the form
$$\exp{\bigl(xz-tH(z)\bigr)}=\sum_{n\ge0}{z^n\over n!}\,h_n(x,t)$$
For the canonical Appell system we have
\begin{equation}\label{eq:canAp}
\exp{\bigl(xz-tH(z)\bigr)}=\sum_{n\ge0}{V(z)^n\over n!}\,y_n(x,t)
\end{equation}
and we recover \eqref{cAppell} via the inversion $z=U(v)$, which we interpret as changing to canonical variables.\bigskip

Observe that each of the polynomials $y_n(x,t)$ is a solution of the evolution equation
$$\frac{\partial u}{\partial t}+H(D)\,u=0$$


\section{Canonical calculus with matrices}
First consider the case $N=1$. We have a function $V(z)$ analytic in a neighborhood of the origin in $\C$, normalized to 
$V(0)=0$, $V'(0)\ne0$. Let $W(z)=1/V'(z)$ have the Taylor expansion
$$W(z)=w_0+w_1z+\cdots+w_kz^k+\cdots$$
The corresponding canonical variable is $Y=XW(D)$, satisfying $[V(D),Y]=\1$. The canonical basis polynomials are
$y_n(x)=Y^n1$, $n\ge0$.\bigskip

Fix the order $p$. Let $\W=W(\D)$. 
Then we employ the algebra generated by the operators $\V=V(\D)$ and $\Y=\X\W$.
Note, e.g., that since $\D^{p+1}=0$, the operators $\V$ and $\W$ are polynomials in $\D$. Similarly, since $\X^{p+1}=0$,
the polynomials $y_n(\X)$ are truncated if $n>p$. However, for $n\le p$, the correspondence between the 
polynomials $y_n(x)$ and vectors $\y_n=y_n(\X)\e_0$ is exact. Namely,  the vector $\y_n$ gives the coefficients of the polynomial
$y_n(x)$. The reason this works is that up to order $p$, the operator $\X$ never acts on a power of $x$ greater than $p$.\\

\subsection{Examples}
\begin{example}\rm
A simple example to illustrate the construction is given by 
$$V(z)=e^z-1\,, \qquad U(v)=\log(1+v)$$
so $W(z)=e^{-z}$, $Y=Xe^{-D}$. 
The relation $X=YU'(V)^{-1}$ reads $X=Y+YV$ or $xy_n=y_{n+1}+ny_n$ yielding the 
recurrence
$$ y_{n+1}=(x-n)y_n$$
for $n>0$.
From $y_0=1$, we easily calculate $$y_n(x)=x(x-1)\cdots(x-n+1)\ .$$
 For $p=4$, with
$\displaystyle\Y=\left( \begin {array}{rrrrr} 0&0&0&0&0\\\noalign{\medskip}1&-1&1&-1&1
\\\noalign{\medskip}0&1&-2&3&-4\\\noalign{\medskip}0&0&1&-3&6
\\\noalign{\medskip}0&0&0&1&-4\end {array} \right) $ we get
$$\Y^2=\left( \begin {array}{rrrrr} 0&0&0&0&0\\\noalign{\medskip}-1&2&-4&8&-
15\\\noalign{\medskip}1&-3&8&-20&43\\\noalign{\medskip}0&1&-5&18&-46
\\\noalign{\medskip}0&0&1&-7&22\end {array} \right), \quad
\Y^3=\left( \begin {array}{rrrrr} 0&0&0&0&0\\\noalign{\medskip}2&-6&18&-53
&126\\\noalign{\medskip}-3&11&-39&130&-327\\\noalign{\medskip}1&-6&29&
-116&313\\\noalign{\medskip}0&1&-9&46&-134\end {array} \right) 
$$

$$\Y^4=\left( \begin {array}{rrrrr} 0&0&0&0&0\\\noalign{\medskip}-6&24&-95&
345&-900\\\noalign{\medskip}11&-50&219&-845&2255\\\noalign{\medskip}-6
&35&-180&754&-2070\\\noalign{\medskip}1&-10&65&-300&849\end {array}
 \right)$$
and 
$$
\Y^5=\left( \begin {array}{rrrrr} 0&0&0&0&0\\\noalign{\medskip}24&-119&559
&-2244&6074\\\noalign{\medskip}-50&269&-1333&5497&-15016
\\\noalign{\medskip}35&-215&1149&-4907&13559\\\noalign{\medskip}-10&75
&-440&1954&-5466\end {array} \right) 
$$
with the first column giving the coefficients of the corresponding polynomial $y_n$,
where, since the leading coefficient equals one, we can see the truncation beginning in this last.\\

\end{example}

\begin{example}\rm
Another interesting example is the Gaussian with drift $\alpha>0$,
$$V(z)=\alpha z-z^2/2\,, \qquad U(v)=\alpha-\sqrt{\alpha^2-2v}$$
the minus sign taken in $U(v)$ to have $U(0)=0$. Then $\displaystyle W(z)=\frac{1}{\alpha-z}$, and
$$\Y=\left( \begin {array}{ccccc} 0&0&0&0&0\\\noalign{\medskip}{\alpha}^{-
1}&{\alpha}^{-2}&2\,{\alpha}^{-3}&6\,{\alpha}^{-4}&24\,{\alpha}^{-5}
\\\noalign{\medskip}0&{\alpha}^{-1}&2\,{\alpha}^{-2}&6\,{\alpha}^{-3}&
24\,{\alpha}^{-4}\\\noalign{\medskip}0&0&{\alpha}^{-1}&3\,{\alpha}^{-2
}&12\,{\alpha}^{-3}\\\noalign{\medskip}0&0&0&{\alpha}^{-1}&4\,{\alpha}
^{-2}\end {array} \right) 
$$
Powers of $\Y$ yield the canonical polynomials, the first few of which are
\begin{eqnarray*}
y_1&=&{\frac {x}{\alpha}}\\
y_2&=&{\frac {x}{{\alpha}^{3}}}+{\frac {{x}^{2}}{{\alpha}^{2}}}\\
y_3&=&3\,{\frac {x}{{\alpha}^{5}}}+3\,{\frac {{x}^{2}}{{\alpha}^{4}}}+{
\frac {{x}^{3}}{{\alpha}^{3}}}\\
y_4&=&15\,{\frac {x}{{\alpha}^{7}}}+15\,{\frac {{x}^{2}}{{\alpha}^{6}}}+6\,{
\frac {{x}^{3}}{{\alpha}^{5}}}+{\frac {{x}^{4}}{{\alpha}^{4}}}\\
y_5&=&105\,{\frac {x}{{\alpha}^{9}}}+105\,{\frac {{x}^{2}}{{\alpha}^{8}}}+45
\,{\frac {{x}^{3}}{{\alpha}^{7}}}+10\,{\frac {{x}^{4}}{{\alpha}^{6}}}+
{\frac {{x}^{5}}{{\alpha}^{5}}}\\
\end{eqnarray*}
These are a scaled variation of Bessel polynomials. \\
In this case 
$\displaystyle U'(V)^{-1}=\alpha\left(1-\frac{2V}{\alpha^2}\right)^{1/2}$. Thus,
expanding and rearranging the relation $X=YU'(V)^{-1}$,
$$\alpha Y=X+\alpha Y\left({\frac {V}{\alpha^2}}+\frac{1}{2}\,{\frac {{V}^{2}}{{\alpha}^{4}}}+\frac{1}{2}\,{
\frac {{V}^{3}}{{\alpha}^{6}}}+\frac{5}{8}\,{\frac {{V}^{4}}{{\alpha}^{8}}}+{
\frac {7}{8}}\,{\frac {{V}^{5}}{{\alpha}^{10}}}+{\frac {21}{16}}\,{
\frac {{V}^{6}}{{\alpha}^{12}}}+{\frac {33}{16}}\,{\frac {{V}^{7}}{{
\alpha}^{14}}}+\ldots\right)$$
which translates to
\begin{eqnarray*}
\alpha\,y_{n+1}&=&xy_n+\frac{n}{\alpha}\,y_{n}+\frac{n(n-1)}{2\alpha^3}\,y_{n-1}+
\frac{n(n-1)(n-2)}{2\alpha^5}\,y_{n-2}+ \ldots\vstrut\\
&=& xy_n+\frac{n}{\alpha}\,y_n+
\sum_{k=2}^n {n\choose k}\,\frac{(2k-3)!!}{\alpha^{2k-1}}\,y_{n-k+1}
\end{eqnarray*}
\end{example}

\begin{example}     \rm
Our final example in this section, involves the LambertW function, which we denote $\cal W$
to avoid confusion with our $W$. Take $V(z)=ze^{-z}$ \cite[v.\th1, p.\th 110]{FS2}. Then $U(v)=-{\cal W}(-v)$.
We find $Y=Xe^D(I-D)^{-1}$ and with $p=7$ the corresponding matrix 
$$\Y=\left( \begin {array}{cccccccc} 0&0&0&0&0&0&0&0\\\noalign{\medskip}1&
2&5&16&65&326&1957&13700\\\noalign{\medskip}0&1&4&15&64&325&1956&13699
\\\noalign{\medskip}0&0&1&6&30&160&975&6846\\\noalign{\medskip}0&0&0&1
&8&50&320&2275\\\noalign{\medskip}0&0&0&0&1&10&75&560
\\\noalign{\medskip}0&0&0&0&0&1&12&105\\\noalign{\medskip}0&0&0&0&0&0&
1&14\end {array} \right) 
$$
One can show that $y_n=x(x+n)^{n-1}$ and that the relation $X=YU'(V)^{-1}$ leads to the recurrence
$$y_{n+1}=(x+2n)y_n+\sum_{k=1}^{n-1}{n\choose k+1}\,k^k\,y_{n-k}\ .$$
\end{example}

\subsection{Matrix expansions}

Using the unit matrices $E_{ij}$ of size $(p+1)\times(p+1)$, we can write formulas for the basic operators. 
We have used in Theorem \ref{thmSL} the expressions
$$ \D=\sum_{k=1}^p k\, E_{k\,k+1} \qquad\hbox {and }\qquad \X=\sum_{k=1}^p  E_{k+1\,k}$$

Then induction yields 
$$\D^j=\sum_{ {1\le k\le p \atop } \atop 1\le k+j\le p+1} (k)_j\, E_{k\,k+j}$$
with $(k)_j=k(k+1)\cdots(k+j-1)$ denoting the rising factorial. Multiplying by $\X$ gives
$$\Y=\sum_{ {1\le k\le p \atop } \atop 1\le k+j\le p+1} E_{k+1\,k+j}\, (k)_j\, w_j $$

For $N>1$, the matrices for $D_j$ and $X_j$ provide the operators
$\Y_j=\X_\mu W_{\mu i}(\D)$ as matrices. Repeated multiplication on the vacuum vector $\e_0$ yields exactly the coefficients
of the polynomials $y_n$ up to order $p$, that is, no variable $x_i$ appears to a power higher than $p$.

\section{Krawtchouk polynomials}

{\it Note.\/} In this section $N$ will indicate a discrete time parameter taking values $N=0,1,2,\ldots$. We will be working
with polynomials in one variable $x$.\bigskip

The Krawtchouk polynomials occur as polynomials orthogonal with respect to the binomial distribution. Here we take
the distribution of the sum of $N$ independent Bernoulli random variables taking values $\pm1$ each with probability $1/2$.
We start with the generating function
\begin{equation}\label{genf}
G(v;x,N)=(1+v)^{(N+x)/2}\,(1-v)^{(N-x)/2}= \sum_{n\ge0}{v^n\over n!}\,K_n(x,N)
\end{equation}
where for $x$ the position of the random walk after $N$ steps, $(N+x)/2$ is the number of positive jumps and
$(N-x)/2$ the number of negative jumps. To see this in the form of a canonical Appell system, write
$$G(v;x,N)=e^{xU(v)-tH(U(v))}=\left(\frac{1+v}{1-v}\right)^{x/2}\,(1-v^2)^{N/2}$$
where we identify $t\leftrightarrow N$,
$$U(v)=\frac12\,\log\frac{1+v}{1-v}\quad\hbox{and\ }\quad H(U(v))=-\frac12\,\log(1-v^2)$$
so that
$$V(z)=\tanh z\quad\hbox{and\ }\quad H(z)=\log\cosh z$$
with $W(z)=1/V'(z)=\cosh^2z$. Now verify that
$$(\cosh D)\,\left({1+v\over1-v}\right)^{x/2}=\frac12\,(e^D+e^{-D})\,\left({1+v\over1-v}\right)^{x/2}=(1-v^2)^{-1/2}\left(\frac{1+v}{1-v}\right)^{x/2}$$
so that
$$G(x,v;N)=e^{-NH(D)}G(x,v;0)=(\sech D)^Ne^{xU(v)}$$
appropriately for the canonical Appell system. \bigskip

The recurrence relation for $\{K_n\}$ is derived as follows. The recurrence relation for the canonical polynomials with $N=0$ is
$X=YU'(V)^{-1}=Y(1-V^2)$. Let $y_n(x)=K_n(x,0)$. Then $K_n(x,N)=(\sech D)^N\,y_n(x)$. The recurrence for $y_n$ is
\begin{equation}\label{sech}
x\,y_n=y_{n+1}-n(n-1)\,y_{n-1}
\end{equation}
Now use the relation $[f(D),x]=f'(D)$ for a function $f$ analytic in a neighborhood of the origin to get
\begin{eqnarray}\label{sechd}
[(\sech D)^N,x]&=&-N(\sech D)^{N-1}(\sech D\,\tanh D)\nonumber \\ &=&-N(\sech D)^N\,\tanh D=-N e^{-NH(D)}V(D)
\end{eqnarray}
thus, applying $(\sech D)^N$ to \eqref{sech} yields
$$x\,K_n-NnK_{n-1}=K_{n+1}-n(n-1)K_{n-1}$$
or
\begin{equation}\label{recc}
x\,K_n=K_{n+1}+n(N-n+1)K_{n-1}
\end{equation}
with $K_{-1}=0$, $K_0=1$.

\subsection{Krawtchouk ghosts}
The raising operator for the polynomials $y_n$, when $N=0$, is
$$Y=XW(D)=x\cosh^2D$$
Since $K_n=(\sech D)^Ny_n$, we have the raising operator
$R$, $RK_n=K_{n+1}$, given by
\begin{eqnarray*}
R&=&(\sech D)^NY(\sech D)^{-N}=(\sech D)^NY(\cosh D)^N\\
&=&(\sech D)^Nx(\cosh D)^{N+2}=x\cosh^2 D-N\sinh D\,\cosh D
\end{eqnarray*}
using \eqref{sechd}. Acting on polynomials, with $V(D)=\tanh D$, we have the commutation relation
$$[V(D),R]=\1$$ 
So we have a representation of the HW-algebra with the commutator equal to the identity.\bigskip

Now notice that if we set $n=N+1$ and $n=N+2$ in \eqref{recc}, we get
$$xK_{N+1}=K_{N+2}\quad\hbox{and}\quad xK_{N+2}=K_{N+3}-(N+2)K_{N+1}$$
so that for $n>N$, all $K_n$ have $K_{N+1}$ as a common factor. Expanding the binomials in \eqref{genf}, we have
$$K_n/n!=\sum_{k\ge0}{(N+x)/2\choose n-k}{(N-x)/2\choose k}\,(-1)^k$$
The random walk can land on any of the points $N, N-2,\ldots,2-N,-N$ in $N$ steps. Let $x=N-2j$, where $0\le j\le N$,
$j$ being the number of negative jumps of the random walk.
Substituting into the generating function \eqref{genf}:
$$G(v;N-2j,N)=(1+v)^{N-j}\,(1-v)^{j}= \sum_{n\ge0}{v^n\over n!}\,K_n(N-2j,N)$$
The left side is a polynomial in $v$ of degree $N$, so that all of the coefficients beyond $N$ vanish identically.
So, $K_n(N-2j,N)=0$ for $0\le j\le N$, $n>N$.
That is, on the support of the binomial distribution,
$K_{N+1}$ vanishes. Thus 
$$0=\|K_{N+1}\|^2=\|K_{N+2}\|^2=\ldots=\|K_{N+k}\|^2=\cdots$$
for $k\ge1$, i.e., the $L^2$-norms of all of the polynomials $K_n$ vanish for $n>N$. So in the Hilbert space, these are zero. From the HW-algebra point of
view, these are thus ``ghost states".\bigskip

On the other hand, if we consider the lowering operator $L$ satisfying 
$$LK_n=n(N-n+1)K_{n-1}$$
We have the commutation relations
$$[L,R]=\Lambda\,,\quad [R,\Lambda]=2R\,,\quad [\Lambda,L]=2L$$
which give a representation of $\?sl?(2)$. 
In fact, for $N\ge0$, we recover the irreducible representations of $\?su?(2)$.\bigskip

\subsection{Krawtchouk calculus with matrices}
In the Krawtchouk Hilbert space, we consider only functions on the spectrum of the operator $X$, namely the finite set of
points $N,N-2,\ldots,2-N,-N$. We have an $(N+1)$-dimensional space spanned by the polynomials $K_n(x,N)$, $0\le n\le N$ of
bounded degree. So in this space, the matrix representations $\hat X$ and $\hat D$ of order $(N+1) \times (N+1)$ will give exact results.\bigskip

Fix an order $p$ and time parameter $N$. Take $p=N$ to get the full basis for time $N$.
First, construct $\hat Y=\hat X\cosh^2\hat D$, using $e^{\pm\hat D}$. Compute $\hat Y^n$, for $0\le n\le p$.
The first column of $\hat Y^n$ are the coefficients of the ``time-zero" polynomial $y_n$.
Let $\hat S_N=(\sech \hat D)^N$, using $\sech \hat D=(\cosh \hat D)^{-1}$.
Then, up to order $p$, for any $N\ge 0$,
$$\hat K_n(N)=\hat S_N\hat Y_n$$
has first column the coefficients of $K_n(x,N)$.\bigskip

\begin{example}\rm
We illustrate with $N=5$. We find
$$\cosh\hat D= \left( \begin {array}{cccccc} 1&0&1&0&1&0\\\noalign{\medskip}0&1&0&3&0
&5\\\noalign{\medskip}0&0&1&0&6&0\\\noalign{\medskip}0&0&0&1&0&10
\\\noalign{\medskip}0&0&0&0&1&0\\\noalign{\medskip}0&0&0&0&0&1
\end {array} \right) $$
and
$$\sech\hat D=  \left( \begin {array}{cccccc} 1&0&-1&0&5&0\\\noalign{\medskip}0&1&0&-
3&0&25\\\noalign{\medskip}0&0&1&0&-6&0\\\noalign{\medskip}0&0&0&1&0&-
10\\\noalign{\medskip}0&0&0&0&1&0\\\noalign{\medskip}0&0&0&0&0&1
\end {array} \right) $$
The raising operator at time-zero is
$$\hat Y=  \left( \begin {array}{cccccc} 0&0&0&0&0&0\\\noalign{\medskip}1&0&2&0&
8&0\\\noalign{\medskip}0&1&0&6&0&40\\\noalign{\medskip}0&0&1&0&12&0
\\\noalign{\medskip}0&0&0&1&0&20\\\noalign{\medskip}0&0&0&0&1&0
\end {array} \right) 
$$
Taking $p=5$, we collect
$$\hat Y^2= \left( \begin {array}{cccccc} 0&0&0&0&0&0\\\noalign{\medskip}0&2&0&20
&0&240\\\noalign{\medskip}1&0&8&0&120&0\\\noalign{\medskip}0&1&0&18&0&
280\\\noalign{\medskip}0&0&1&0&32&0\\\noalign{\medskip}0&0&0&1&0&20
\end {array} \right) ,\,
\hat Y^3= \left( \begin {array}{cccccc} 0&0&0&0&0&0\\\noalign{\medskip}2&0&24&0
&496&0\\\noalign{\medskip}0&8&0&168&0&2720\\\noalign{\medskip}1&0&20&0
&504&0\\\noalign{\medskip}0&1&0&38&0&680\\\noalign{\medskip}0&0&1&0&32
&0\end {array} \right) $$
$$\hat Y^4= \left( \begin {array}{cccccc} 0&0&0&0&0&0\\\noalign{\medskip}0&24&0&
640&0&10880\\\noalign{\medskip}8&0&184&0&4800&0\\\noalign{\medskip}0&
20&0&624&0&10880\\\noalign{\medskip}1&0&40&0&1144&0
\\\noalign{\medskip}0&1&0&38&0&680\end {array} \right) ,\,
\hat Y^5= \left( \begin {array}{cccccc} 0&0&0&0&0&0\\\noalign{\medskip}24&0&688
&0&18752&0\\\noalign{\medskip}0&184&0&5904&0&103360
\\\noalign{\medskip}20&0&664&0&18528&0\\\noalign{\medskip}0&40&0&1384&0
&24480\\\noalign{\medskip}1&0&40&0&1144&0\end {array} \right) $$
corresponding to the time-zero polynomials $y_n$, $0\le n\le5$. These work for any $N$. For example, for $K_4(x,3)$, compute
$$(\sech\hat D)^3\hat Y^4= \left( \begin {array}{cccccc} 9&0&768&0&23352&0\\\noalign{\medskip}0&
9&0&1294&0&25160\\\noalign{\medskip}-10&0&-536&0&-15792&0
\\\noalign{\medskip}0&-10&0&-516&0&-9520\\\noalign{\medskip}1&0&40&0&
1144&0\\\noalign{\medskip}0&1&0&38&0&680\end {array} \right) $$
or $K_4(x,3)=x^4-10x^2+9=(x^2-1^2)(x^2-3^2)$, vanishing at $x=\pm1,\pm3$.\bigskip

And $K_4(x,5)$ corresponds to
$$(\sech\hat D)^5\hat Y^4=  \left( \begin {array}{cccccc} 0&2480&0&88120&0&1564000
\\\noalign{\medskip}149&0&7728&0&227032&0\\\noalign{\medskip}0&-1016&0
&-35616&0&-631040\\\noalign{\medskip}-30&0&-1336&0&-38672&0
\\\noalign{\medskip}0&40&0&1384&0&24480\\\noalign{\medskip}1&0&40&0&
1144&0\end {array} \right) 
 $$
Note that, in fact, one only needs the first column of $\hat Y^n$ for each $n$, then applying appropriate powers of
$\sech \hat D$ produces the polynomials for $N>0$.\bigskip

Here are the polynomials for $N=5$:
$$
\begin{array}{lll}
&K_0=1,   &\quad K_1= x, \\
&K_2=x^2-5,  &\quad K_3= x^3-13x, \\
&K_4=x^4-22x^2+45, &\quad K_5=x^5-30x^3+149x 
\end{array}
$$
and 
$$K_6=x^6-35x^4+259x^2-225=(x^2-1^2)(x^2-3^2)(x^2-5^2)$$
accordingly.
\end{example}

\subsection{Krawtchouk expansions}
The matrix method can be used to compute Krawtchouk expansions for functions $f$ defined on the spectrum
$-N,2-N,\ldots,N-2,N$ for $N>0$. Start with 
$$G(v;x,N)=\left(\frac{1+v}{1-v}\right)^{x/2}\,(1-v^2)^{N/2}=\sum_{n\ge0}{v^n\over n!}\,K_n(x,N)$$
Substitute $v=V(z)=\tanh z$ and rearrange to get, cf. \eqref{eq:canAp},
$$e^{zx}=(\cosh z)^N\sum_{n\ge0}{(\tanh z)^n\over n!}\,K_n(x,N)$$
Replacing $z$ by $D_s=d/ds$, apply both sides to a function $f(s)$:
$$e^{xD_s}f(s)=f(s+x)=(\cosh D_s)^N\sum_{n\ge0}{(\tanh D_s)^n\over n!}\,K_n(x,N)f(s)$$
Letting $s=0$, thinking of $f$ as a function of $x$ instead of $s$,
we can replace $D_s$ by our usual $D=d/dx$,  to get
$$f(x)=\sum_{0\le n\le N}{K_n(x,N)\over n!}(\cosh D)^N(\tanh D)^nf(0)$$
In other words, the coefficients of the Krawtchouk expansion of $f(x)$ are given by
$$\tilde f(n)=\frac{1}{n!}\,(\cosh D)^N(\tanh D)^nf(0)=\frac{1}{n!}\,(\cosh D)^{N-n}(\sinh D)^nf(0)$$
If $f(x)$ is a polynomial in $x$, we can construct a vector from its coefficients, apply the matrices
$(\cosh \hat D)^N(\tanh \hat D)^n$, rescaling by $n!$, and find its Krawtchouk expansion. 

\begin{example}\rm
Here are the matrices $(\cosh\hat D)^5(\tanh\hat D)^n/n!$ for $N=5$, \hfill\break $0\le n\le 5$. 
For $n=0,1,2$:
$$
\left( \begin {array}{cccccc} 1&0&5&0&65&0\\\noalign{\medskip}0&1&0&
15&0&325\\\noalign{\medskip}0&0&1&0&30&0\\\noalign{\medskip}0&0&0&1&0&
50\\\noalign{\medskip}0&0&0&0&1&0\\\noalign{\medskip}0&0&0&0&0&1
\end {array} \right) ,\,
 \left( \begin {array}{cccccc} 0&1&0&13&0&241\\\noalign{\medskip}0&0&2
&0&52&0\\\noalign{\medskip}0&0&0&3&0&130\\\noalign{\medskip}0&0&0&0&4&0
\\\noalign{\medskip}0&0&0&0&0&5\\\noalign{\medskip}0&0&0&0&0&0
\end {array} \right) ,\,
 \left( \begin {array}{cccccc} 0&0&1&0&22&0\\\noalign{\medskip}0&0&0&3
&0&110\\\noalign{\medskip}0&0&0&0&6&0\\\noalign{\medskip}0&0&0&0&0&10
\\\noalign{\medskip}0&0&0&0&0&0\\\noalign{\medskip}0&0&0&0&0&0
\end {array} \right) 
$$
For $n=3,4,5$:
$$
 \left( \begin {array}{cccccc} 0&0&0&1&0&30\\\noalign{\medskip}0&0&0&0
&4&0\\\noalign{\medskip}0&0&0&0&0&10\\\noalign{\medskip}0&0&0&0&0&0
\\\noalign{\medskip}0&0&0&0&0&0\\\noalign{\medskip}0&0&0&0&0&0
\end {array} \right) ,\,
 \left( \begin {array}{cccccc} 0&0&0&0&1&0\\\noalign{\medskip}0&0&0&0
&0&5\\\noalign{\medskip}0&0&0&0&0&0\\\noalign{\medskip}0&0&0&0&0&0
\\\noalign{\medskip}0&0&0&0&0&0\\\noalign{\medskip}0&0&0&0&0&0
\end {array} \right) ,\,
 \left( \begin {array}{cccccc} 0&0&0&0&0&1\\\noalign{\medskip}0&0&0&0
&0&0\\\noalign{\medskip}0&0&0&0&0&0\\\noalign{\medskip}0&0&0&0&0&0
\\\noalign{\medskip}0&0&0&0&0&0\\\noalign{\medskip}0&0&0&0&0&0
\end {array} \right) $$
Let $f(x)=x^4+2x^3-x^2+5x$. One checks by hand, with $N=5$, that
$$f=K_4+2K_3+21K_2+31K_1+60$$
Form the column vector, transpose of $[0,5,-1,2,1,0]$ and apply each of the above matrices. Stacking these as row vectors we have the matrix
$$ \left( \begin {array}{cccccc} 60&35&29&2&1&0\\\noalign{\medskip}31&50
&6&4&0&0\\\noalign{\medskip}21&6&6&0&0&0\\\noalign{\medskip}2&4&0&0&0&0
\\\noalign{\medskip}1&0&0&0&0&0\\\noalign{\medskip}0&0&0&0&0&0
\end {array} \right) $$
whose first column gives the Krawtchouk coefficients.\bigskip

Alternatively, start with the top row of $(\cosh \hat D)^N$. Call it $y_0$. Generate recursively, for $1\le n\le N$,
$$y_{n}=y_{n-1}\,\tanh\D/n$$
and construct the matrix $Y$ with rows $y_0,\ldots,y_n,\ldots,y_N$. Applying to the vector of coefficients of a polynomial
gives the coefficients of the Krawtchouk expansion of that polynomial. In particular,
applying to the vector of coefficients of a Krawtchouk polynomial gives a standard basis vector, zero except for a single entry equal to one. 
That is, $Y^{-1}$ is a matrix whose columns
are the coefficients of the corresponding Krawtchouk polynomials. For the example above, we have
$$Y=\left( \begin {array}{cccccc} 1&0&5&0&65&0\\\noalign{\medskip}0&1&0&13&0&241\\\noalign{\medskip}0&0&1&0&22&0\\\noalign{\medskip}0&0&0&1&0&
30\\\noalign{\medskip}0&0&0&0&1&0\\\noalign{\medskip}0&0&0&0&0&1
\end {array} \right) 
$$
and 
$$Y^{-1}= \left( \begin {array}{cccccc} 1&0&-5&0&45&0\\\noalign{\medskip}0&1&0&
-13&0&149\\\noalign{\medskip}0&0&1&0&-22&0\\\noalign{\medskip}0&0&0&1&0
&-30\\\noalign{\medskip}0&0&0&0&1&0\\\noalign{\medskip}0&0&0&0&0&1
\end {array} \right) 
$$ 
\end{example}

\section{Summary and prospects}
In this article, the one-variable case of the matrix approach has been presented in some detail, along with the
basic theory for the multivariate case. The TAA algebra conveniently replaces the HW algebra in the 
finite-dimensional setting. The connection with orthofermions is interesting and clarifies the underlying
structure. \bigskip

The approach here allows for doing calculations in finite, exact terms. There are many possible applications
including image analysis, optics, as well as quantum information.

\end{document}